\documentclass[reqno,12pt,a4paper]{amsart}

\usepackage{mathrsfs}
\usepackage{amssymb}
\usepackage{amsfonts}
\usepackage{latexsym}
\usepackage{amsthm}

\usepackage{graphicx}
\def\lb{\label}

\newcommand{\er}[1]{\textrm{(\ref{#1})}}

\begin{document}


\renewcommand{\theequation}{\arabic{section}.\arabic{equation}}
\theoremstyle{plain}
\newtheorem{theorem}{\bf Theorem}[section]
\newtheorem{lemma}[theorem]{\bf Lemma}
\newtheorem{corollary}[theorem]{\bf Corollary}
\newtheorem{proposition}[theorem]{\bf Proposition}
\newtheorem{definition}[theorem]{\bf Definition}
\newtheorem{remark}[theorem]{\it Remark}

\def\a{\alpha}  \def\cA{{\mathcal A}}     \def\bA{{\bf A}}  \def\mA{{\mathscr A}}
\def\b{\beta}   \def\cB{{\mathcal B}}     \def\bB{{\bf B}}  \def\mB{{\mathscr B}}
\def\g{\gamma}  \def\cC{{\mathcal C}}     \def\bC{{\bf C}}  \def\mC{{\mathscr C}}
\def\G{\Gamma}  \def\cD{{\mathcal D}}     \def\bD{{\bf D}}  \def\mD{{\mathscr D}}
\def\d{\delta}  \def\cE{{\mathcal E}}     \def\bE{{\bf E}}  \def\mE{{\mathscr E}}
\def\D{\Delta}  \def\cF{{\mathcal F}}     \def\bF{{\bf F}}  \def\mF{{\mathscr F}}
\def\c{\chi}    \def\cG{{\mathcal G}}     \def\bG{{\bf G}}  \def\mG{{\mathscr G}}
\def\z{\zeta}   \def\cH{{\mathcal H}}     \def\bH{{\bf H}}  \def\mH{{\mathscr H}}
\def\e{\eta}    \def\cI{{\mathcal I}}     \def\bI{{\bf I}}  \def\mI{{\mathscr I}}
\def\p{\psi}    \def\cJ{{\mathcal J}}     \def\bJ{{\bf J}}  \def\mJ{{\mathscr J}}
\def\vT{\Theta} \def\cK{{\mathcal K}}     \def\bK{{\bf K}}  \def\mK{{\mathscr K}}
\def\k{\kappa}  \def\cL{{\mathcal L}}     \def\bL{{\bf L}}  \def\mL{{\mathscr L}}
\def\l{\lambda} \def\cM{{\mathcal M}}     \def\bM{{\bf M}}  \def\mM{{\mathscr M}}
\def\L{\Lambda} \def\cN{{\mathcal N}}     \def\bN{{\bf N}}  \def\mN{{\mathscr N}}
\def\m{\mu}     \def\cO{{\mathcal O}}     \def\bO{{\bf O}}  \def\mO{{\mathscr O}}
\def\n{\nu}     \def\cP{{\mathcal P}}     \def\bP{{\bf P}}  \def\mP{{\mathscr P}}
\def\r{\rho}    \def\cQ{{\mathcal Q}}     \def\bQ{{\bf Q}}  \def\mQ{{\mathscr Q}}
\def\s{\sigma}  \def\cR{{\mathcal R}}     \def\bR{{\bf R}}  \def\mR{{\mathscr R}}
\def\S{\Sigma}  \def\cS{{\mathcal S}}     \def\bS{{\bf S}}  \def\mS{{\mathscr S}}
\def\t{\tau}    \def\cT{{\mathcal T}}     \def\bT{{\bf T}}  \def\mT{{\mathscr T}}
\def\f{\phi}    \def\cU{{\mathcal U}}     \def\bU{{\bf U}}  \def\mU{{\mathscr U}}
\def\F{\Phi}    \def\cV{{\mathcal V}}     \def\bV{{\bf V}}  \def\mV{{\mathscr V}}
\def\P{\Psi}    \def\cW{{\mathcal W}}     \def\bW{{\bf W}}  \def\mW{{\mathscr W}}
\def\o{\omega}  \def\cX{{\mathcal X}}     \def\bX{{\bf X}}  \def\mX{{\mathscr X}}
\def\x{\xi}     \def\cY{{\mathcal Y}}     \def\bY{{\bf Y}}  \def\mY{{\mathscr Y}}
\def\X{\Xi}     \def\cZ{{\mathcal Z}}     \def\bZ{{\bf Z}}  \def\mZ{{\mathscr Z}}
\def\O{\Omega}

\newcommand{\mc}{\mathscr {c}}

\newcommand{\gA}{\mathfrak{A}}          \newcommand{\ga}{\mathfrak{a}}
\newcommand{\gB}{\mathfrak{B}}          \newcommand{\gb}{\mathfrak{b}}
\newcommand{\gC}{\mathfrak{C}}          \newcommand{\gc}{\mathfrak{c}}
\newcommand{\gD}{\mathfrak{D}}          \newcommand{\gd}{\mathfrak{d}}
\newcommand{\gE}{\mathfrak{E}}
\newcommand{\gF}{\mathfrak{F}}           \newcommand{\gf}{\mathfrak{f}}
\newcommand{\gG}{\mathfrak{G}}           
\newcommand{\gH}{\mathfrak{H}}           \newcommand{\gh}{\mathfrak{h}}
\newcommand{\gI}{\mathfrak{I}}           \newcommand{\gi}{\mathfrak{i}}
\newcommand{\gJ}{\mathfrak{J}}           \newcommand{\gj}{\mathfrak{j}}
\newcommand{\gK}{\mathfrak{K}}            \newcommand{\gk}{\mathfrak{k}}
\newcommand{\gL}{\mathfrak{L}}            \newcommand{\gl}{\mathfrak{l}}
\newcommand{\gM}{\mathfrak{M}}            \newcommand{\gm}{\mathfrak{m}}
\newcommand{\gN}{\mathfrak{N}}            \newcommand{\gn}{\mathfrak{n}}
\newcommand{\gO}{\mathfrak{O}}
\newcommand{\gP}{\mathfrak{P}}             \newcommand{\gp}{\mathfrak{p}}
\newcommand{\gQ}{\mathfrak{Q}}             \newcommand{\gq}{\mathfrak{q}}
\newcommand{\gR}{\mathfrak{R}}             \newcommand{\gr}{\mathfrak{r}}
\newcommand{\gS}{\mathfrak{S}}              \newcommand{\gs}{\mathfrak{s}}
\newcommand{\gT}{\mathfrak{T}}             \newcommand{\gt}{\mathfrak{t}}
\newcommand{\gU}{\mathfrak{U}}             \newcommand{\gu}{\mathfrak{u}}
\newcommand{\gV}{\mathfrak{V}}             \newcommand{\gv}{\mathfrak{v}}
\newcommand{\gW}{\mathfrak{W}}             \newcommand{\gw}{\mathfrak{w}}
\newcommand{\gX}{\mathfrak{X}}               \newcommand{\gx}{\mathfrak{x}}
\newcommand{\gY}{\mathfrak{Y}}              \newcommand{\gy}{\mathfrak{y}}
\newcommand{\gZ}{\mathfrak{Z}}             \newcommand{\gz}{\mathfrak{z}}

\def\ve{\varepsilon}   \def\vt{\vartheta}    \def\vp{\varphi}    \def\vk{\varkappa}

\def\A{{\mathbb A}} \def\B{{\mathbb B}} \def\C{{\mathbb C}}
\def\dD{{\mathbb D}} \def\E{{\mathbb E}} \def\dF{{\mathbb F}} \def\dG{{\mathbb G}} \def\H{{\mathbb H}}\def\I{{\mathbb I}} \def\J{{\mathbb J}} \def\K{{\mathbb K}} \def\dL{{\mathbb L}}\def\M{{\mathbb M}} \def\N{{\mathbb N}} \def\dO{{\mathbb O}} \def\dP{{\mathbb P}} \def\R{{\mathbb R}}\def\S{{\mathbb S}} \def\T{{\mathbb T}} \def\U{{\mathbb U}} \def\V{{\mathbb V}}\def\W{{\mathbb W}} \def\X{{\mathbb X}} \def\Y{{\mathbb Y}} \def\Z{{\mathbb Z}}


\def\la{\leftarrow}              \def\ra{\rightarrow}            \def\Ra{\Rightarrow}
\def\ua{\uparrow}                \def\da{\downarrow}
\def\lra{\leftrightarrow}        \def\Lra{\Leftrightarrow}


\def\lt{\biggl}                  \def\rt{\biggr}
\def\ol{\overline}               \def\wt{\widetilde}
\def\no{\noindent}


\let\ge\geqslant                 \let\le\leqslant
\def\lan{\langle}                \def\ran{\rangle}
\def\/{\over}                    \def\iy{\infty}
\def\sm{\setminus}               \def\es{\emptyset}
\def\ss{\subset}                 \def\ts{\times}
\def\pa{\partial}                \def\os{\oplus}
\def\om{\ominus}                 \def\ev{\equiv}
\def\iint{\int\!\!\!\int}        \def\iintt{\mathop{\int\!\!\int\!\!\dots\!\!\int}\limits}
\def\el2{\ell^{\,2}}             \def\1{1\!\!1}
\def\sh{\sharp}
\def\wh{\widehat}
\def\bs{\backslash}
\def\intl{\int\limits}

\def\na{\mathop{\mathrm{\nabla}}\nolimits}
\def\sh{\mathop{\mathrm{sh}}\nolimits}
\def\ch{\mathop{\mathrm{ch}}\nolimits}
\def\where{\mathop{\mathrm{where}}\nolimits}
\def\all{\mathop{\mathrm{all}}\nolimits}
\def\as{\mathop{\mathrm{as}}\nolimits}
\def\Area{\mathop{\mathrm{Area}}\nolimits}
\def\arg{\mathop{\mathrm{arg}}\nolimits}
\def\const{\mathop{\mathrm{const}}\nolimits}
\def\det{\mathop{\mathrm{det}}\nolimits}
\def\diag{\mathop{\mathrm{diag}}\nolimits}
\def\diam{\mathop{\mathrm{diam}}\nolimits}
\def\dim{\mathop{\mathrm{dim}}\nolimits}
\def\dist{\mathop{\mathrm{dist}}\nolimits}
\def\Im{\mathop{\mathrm{Im}}\nolimits}
\def\Iso{\mathop{\mathrm{Iso}}\nolimits}
\def\Ker{\mathop{\mathrm{Ker}}\nolimits}
\def\Lip{\mathop{\mathrm{Lip}}\nolimits}
\def\rank{\mathop{\mathrm{rank}}\limits}
\def\Ran{\mathop{\mathrm{Ran}}\nolimits}
\def\Re{\mathop{\mathrm{Re}}\nolimits}
\def\Res{\mathop{\mathrm{Res}}\nolimits}
\def\res{\mathop{\mathrm{res}}\limits}
\def\sign{\mathop{\mathrm{sign}}\nolimits}
\def\span{\mathop{\mathrm{span}}\nolimits}
\def\supp{\mathop{\mathrm{supp}}\nolimits}
\def\Tr{\mathop{\mathrm{Tr}}\nolimits}
\def\BBox{\hspace{1mm}\vrule height6pt width5.5pt depth0pt \hspace{6pt}}


\newcommand\nh[2]{\widehat{#1}\vphantom{#1}^{(#2)}}
\def\dia{\diamond}

\def\Oplus{\bigoplus\nolimits}



\def\qqq{\qquad}
\def\qq{\quad}
\let\ge\geqslant
\let\le\leqslant
\let\geq\geqslant
\let\leq\leqslant
\newcommand{\ca}{\begin{cases}}
\newcommand{\ac}{\end{cases}}
\newcommand{\ma}{\begin{pmatrix}}
\newcommand{\am}{\end{pmatrix}}
\renewcommand{\[}{\begin{equation}}
\renewcommand{\]}{\end{equation}}
\def\eq{\begin{equation}}
\def\qe{\end{equation}}
\def\[{\begin{equation}}
\def\bu{\bullet}

\title[{Trace formulas for Schr\"odinger operators with complex potentials}]
{Trace formulas for Schr\"odinger operators with complex potentials
on half-line}

\date{\today}

\author[Evgeny Korotyaev]{Evgeny Korotyaev}
\address{Department of Math. Analysis, Saint-Petersburg State University,
Universitetskaya nab. 7/9, St. Petersburg, 199034, Russia, \ E-mail
address: korotyaev@gmail.com, \ e.korotyaev@spbu.ru }

\subjclass{} \keywords{Complex potentials, Trace formulas}

\begin{abstract}
 We consider Schr\"odinger operators with complex-valued
decreasing potentials  on the half-line. Such operator  has
essential spectrum on the half-line plus eigenvalues (counted with
algebraic multiplicity) in the complex plane without the positive
half-line. We determine trace formula: sum of Im part of these
eigenvalues plus some singular measure plus some integral  from the
Jost function. Moreover, we estimate of sum of Im part of
eigenvalues and singular measure in terms of the norm of potentials.
In addition, we get bounds on the total number of eigenvalues, when
the potential is compactly supported.
\end{abstract}

\maketitle


\section {Introduction and main results}
\setcounter{equation}{0}

\subsection{Introduction}
We consider the Schr\"odinger operator $Hf=-f''+qf, \  f(0)=0$ on
$L^2(\R_+ )$. We assume that the potential $q$  is  complex and
satisfies:
\[
\lb{dV} \int_0^\iy(1+x)|q(x)|dx<\iy.
\]
It is well-known that the operator $H$ has essential  spectrum
$[0,\iy)$  plus $N\in [0, \iy]$ eigenvalues (counted with
multiplicity) in the cut domain $\C\sm [0,\iy)$. The eigenvalues
(counted with multiplicity) of the operator $H$ in the cut domain
$\C\sm [0,\iy)$ we denote by $E_j, j=1,...,N$. Note, that the
multiplicity of each eigenvalue equals 1, but we call the
multiplicity of the eigenvalue its algebraic multiplicity. Instead
of the energy $E\in \C\sm \R_+$ we will use the momentum defined by
$k=\sqrt E\in \ol\C_+$, where $\C_\pm=\{\pm \Im z>0\}$. We call
$k_j=\sqrt E_j\in\C_+$ also the eigenvalues of the operator $H$. Of
course, $E$ is really the energy, but since $k$ is the natural
parameter, we will abuse terminology. We assume that
$k_1,..,k_{N}\in \C_+$ are labeled by
\[
\lb{Lkn} \Im k_1\ge \Im k_2\ge \Im k_3\ge... \ge \Im k_n\ge ...
\]

We shortly describe results about trace formulas:

$\bu $ In 1960 Buslaev and Faddev  \cite{BF60} determined the
classical results about trace formulas for Schr\"odinger operators
with real decaing potentials on half-line.

$\bu $ The  multidimensional case was studied in \cite{B66}. Trace
formulas for Stark operators and magnetic Schr\"odinger operators
were discussed in \cite{KP03}, \cite{KP04}.

$\bu $ The trace formulas for Schr\"odinger operators with real
periodic potentials are determined in \cite{KK95,K97}. They were
used to obtain two-sided estimates of potential in terms of gap
lengths (or the action variables for KdV) in \cite{K00} via the
conformal mapping theory for the quasimomentum.

$\bu $  Trace formulas for Schr\"odinger operators with complex
potentials were considered  on the lattice $\Z^d$ (see \cite{K17},
\cite{KL18}, \cite{MN15}) and on $\R^3$  \cite{K17x}.

Our main goal is to determine trace formulas for Schr\"odinger
operators with complex potentials on the half-line. Our trace
formula is the identity \er{tr1},  where the left hand side is the
integral from the real part of potential, the sum of $\Im k_j$ and
the integral of the singular measure and the right hand side is the
integral from $\log |\p(k+i0)\p(-k+i0)|$ on the positive half-line.
Here we have the new term, the singular measure, which is absent for
real potentials. However, in \er{eBs} we estimates the singular
measure and the sum of $\Im k_j$   in terms of the potential. In our
consideration the technique from \cite{K17x} about Schr\"odinger
operators with complex-valued potentials on $\R^3$, where  the Hardy
spaces in the upper half-space were used. Note that in the case of
lattice \cite{K17}, \cite{KL18} the Hardy spaces on the unit disc
were used.

In contrast to the trace formula for complex potentials, there are
many results on estimates of eigenvalues in terms of potentials, see
recent articles \cite{F18, FS17} and references therein.
 In our 1-dimensional case  there exist
many results about bounds on sums of powers of eigenvalues of
Schr\"odinger operators with complex-valued potentials in terms of
$L^p$-norms of the potentials see \cite{DHK09, H11, LS09, S10, Sa10}
and references therein.

\subsection{The Jost function and the Hardy spaces}
We recall the well-known facts about the Schr\"odinger operator $H$,
see e.g., \cite{F59}.  Introduce the Jost solutions $f_+(x,k )$  of
the equation
\[
\lb{1.3} -f_+''+qf_+=k ^2f_+,\ \ \ x\geq 0, \ \ \ k \in \ol\C_+\sm
\{0\} ,
\]
with the conditions $f_+(x,k )=e^{ixk }+o(1)$ as $ x\to \iy$ and $
k\in \R\sm \{0\}$.
 We define the Jost function $\p(k)=f_+(0,k )$.
The Jost function $\p$ is analytic in  $\C_+ $,  continuous up to
the real line and satisfies
\[
\lb{asp1V} \p(k)=1-{Q_0+o(1)\/ik}\qqq \as \qqq |k |\to \iy,
\]
uniformly in $\arg k \in [0,\pi]$, where $Q_0={1\/2}\int_0^\iy
q(t)dt $. The function $\p$ has $N\in [0,\iy] $ zeros in $\C_+$
given by $k_{j}=\sqrt{E_j}\in \C_+$, counted with algebraic
multiplicity.  Define the set $\s_d=\{k_1,...,k_{N}\in \C_+\}$. Due
to \er{asp1V} the set  $\s_d$ is bounded and satisfies  (see
\cite{FLS11})
\[
\lb{1.2}
\begin{aligned}
\s_d \ss  \{k\in \ol\C_+: \p(k)=0\}     \ss
\{k\in \ol\C_+: |k|\le r_{c}\},\qq  r_{c}=\|q\|,  \\
 \where\qq  \|q\|_\a=\int_0^\iy  x^\a |q(x)|dx, \ \a\ge 0, \qqq \|q\|=\|q\|_0.
 \end{aligned}
\]

Let a function $F(k), k=u+iv\in \C_+$ be analytic on $\C_+$. For
$0<p\le \iy$ we say $F$ belongs the Hardy space $ \mH_p=\mH_p(C_+)$
if $F$ satisfies $\|F\|_{\mH_p}<\iy$, where $\|F\|_{\mH_p}$ is given
by
$$
\|F\|_{\mH_p}=\ca
\sup_{v>0}{1\/2\pi}\rt(\int_\R|F(u+iv))|^pdu\rt)^{1\/p} &
if  \qqq 0< p<\iy\\
 \sup_{k\in \C_+}|F(k)| & if \qqq p=\iy\ac .
$$
Note that the definition of the Hardy space $\mH_p$ involves all
$v=\Im k>0$.

We remark  that the Jost function $\p\in \mH_\iy$, since $\p$ is
uniformly bounded in $\C_+$. Due to \er{1.2} all zeros of $\p$ are
uniformly bounded, then we can define the Blaschke product $B$ by
\begin{equation}
\label{Bk}
 B(k)=\prod_{j=1}^N\rt(\frac{k-k_j}{ k-\ol k_j}\rt),\qqq k\in \C_+.
\end{equation}
 We describe the basic properties
of the Blaschke product $B$ as an analytic function in $\C_+$.

\begin{proposition}\lb{T1}
Let a potential $q$ satisfy \er{dV}. Then $\p\in \mH_\iy$ and is
continuous  up to the real line and satisfies
\[
\lb{J1} |\p(k)|\le e^{\o},\qqq \forall \ \o\in
\rt\{\|q\|_1,{\|q\|\/|k|}\rt\},
\]
and if $\|q\|_1<\o_*$, where $\o_*e^{\o_*}=1$, then the operator $H$
has not eigenvalues.

i) The zeros $(k_j)_1^N$ of $\p$  in the upper-half plane $\C_+$
(counted with multiplicity) labeled by \er{Lkn} satisfy
\[
\lb{B1} \sum _{j=1}^N \Im k_j<\iy.
\]

The Blaschke product $B(k)$ given by \er{Bk} converges absolutely
and  uniformly in every bounded disc in $\C_+\sm \s_d$, and the
function $B\in \mH_\iy$ with $\|B\|_{\mH_\iy}\le 1$.

 ii) The Blaschke product $B$ has an analytic continuation from $\C_+$
into the domain $\{|k|>r_c\}$, where $r_c=\|q\|$ and has the
following Taylor series
\[
\begin{aligned}
\lb{B6}
\log  B(z)=-i{B_0\/k}-i{B_1\/2k^2}-i{B_2\/3k^3}-..., \qqq as \qqq |k|>r_c,\\
 B_0=2\sum_{j=1}^N\Im k_j,\qqq B_n=2\sum_{j=1}^N\Im k_j^{n+1},\qq n\ge 1,
\end{aligned}
\]
where each sum $B_n,  n\ge 1$ is absolutely convergence and
satisfies
\[
|B_n|\le 2\sum_{j=1}^N|\Im k_j^{n+1}|\le {\pi}(n+1)r_c^{n}B_0.
\]

\end{proposition}

\no {\bf Remark.} 1) The function $B$ has a complicated properties
in the disk $\{|k|<r_c\}, r_c=\|q\|$ and very good properties in the
domain $\{|k|>r_c\}$.

2) We use asymptotics of $B$ at large $|k|$  to determine the trace
formulas in Theorem \ref{T3}.

\no {\bf Example.} Consider the potential $q(x)=At x^{t^2-1}, x\in
(0,1), t>0$ and $q(x)=0$ for $x>1$, where $A\in \C$. We have
$\|q\|=|A|/t$ and $\|q\|_1={|A|t\/1+t^2}$. If $t$ is small, then the
complex potential $q$ is rather big and all eigenvalues belong to
the half-disk with the radius $\|q\|=|A|/t$, but if  $\|q\|_1<\o_*$,
where $\o_*e^{\o_*}=1$, then by Proposition \ref{T1}, the operator
$H$ has not eigenvalues.

\subsection{Trace formulas and estimates}
We describe the Jost function $\p\in \mH_\iy$ in terms of a
canonical factorization.

\begin{theorem}\label{T2}
Let a potential $q$ satisfy \er{dV}. Then $\p$ has a canonical
factorization in $\C_+$ given by
\[
\lb{Dio} \p=\p_{in}\p_{out},
\]
where

$\bu$ $\p_{in}$ is the inner factor of $\p$ having the form
\[
\lb{Di1} \p_{in}(k)=B(k)e^{-iK(k)},\qqq  K(k)={1\/\pi}\int_\R
{d\n(t)\/k-t}\qqq \forall \ k\in \C_+.
\]
$\bu$  $B$ is the Blaschke product for $\Im k>0$ given by \er{Bk}
and $d\n(t)\ge 0$ is some singular compactly supported measure on
$\R$, which satisfies
\[
\lb{smsx}
\begin{aligned}
\n(\R)=\int_\R d\n(t)<\iy,\\
 \supp \n \ss \{z\in \R: \p(z)=0\}\ss [-r_c, r_c].
 \end{aligned}
\]
$\bu$  The function $K(\cdot)$ has an analytic continuation from
$\C_+$ into the cut domain $\C\sm [-r_c, r_c]$   and has the
following Taylor series in the domain $\{|k|>r_c\}$:
\[
\lb{Kn}
\begin{aligned}
 K(k)=\sum_{j=0}^\iy {K_j\/k^{j+1}},\qqq  \qq K_j={1\/\pi}\int_\R
t^jd\n(t).
 \end{aligned}
\]
 $\bu$ $\p_{out}$ is the outer factor given by
\[
\lb{Do2} \p_{out}(k)=e^{iM(k)},\qqq  M(k)= {1\/\pi}\int_\R {\log
|\p(t)|\/k-t} dt,\qq k\in \C_+,
\]
where the function $\log |\p(t+i0)|$ belongs to  $L_{loc}^1(\R)$.

\end{theorem}

\no {\bf Remark.} 1)
 Due to \er{asp1V} the integral $M(k)$ in \er{Do2} converges
absolutely.

\no 2) We have $\p_{in}\le 1$, since $B\le 1$ and $\Im {1\/k-t}\le
0$ for all $k\in \C_+$.

\no 3) These results are crucial to determine trace formulas in
Theorem \ref{T3}. The canonical factorization is a first trace
formula. It is a generating function, these results will be used in
the proof of trace formulas in Theorem \ref{T3}.

\medskip

Let $R(k)=(H-k^2)^{-1}$ and $R_0(k)=(H_0-k^2)^{-1}$, where $H_0$ is
the operator $H$ at $q=0$. The differentiation of a canonical
factorization produces a trace formula for $\Tr (R(k)-R_0(k))$.

\begin{corollary}
\lb{T2x} Let a potential $q$ satisfy \er{dV}. Then  the trace
formula
\[
\lb{tre1} -2k\Tr \rt(R(k)-R_0(k) \rt)= \sum {2i\Im
k_j\/(k-k_j)(k-\ol k_j)}+{i\/\pi}\int_{\R}{d\m(t)\/(t-k)^2},
\]
holds true for any $k\in \C_+\sm\s_d$, where  the measure
$d\m(t)=\log |\p(t)|dt-d\n(t)$ and the series converges uniformly in
every bounded disc in $\C_+\sm \s_d$.

\end{corollary}

We recall the well-known results about the asymptotics of the Jost
function. Introduce the Sobolev space $W_m$ given by
\[
W_m=\lt\{xq, q\in L^1(\R_+): q^{(j)}\in L^1(0,\iy),\  j=1,..,
m\lt\},\qq m\geq 0.
\]
If $q\in W_{m+1}, m\ge 0$, then the function $\p(\cdot)$ is analytic
in $\C_+$ and continuous up to the real line and  satisfies
\[
\lb{apm}
  i\log \p(k)=-{Q_0\/k }-{Q_1\/k ^{2}}-{Q_2\/k ^{3}}+\dots
  -{Q_m+o(1)\/k ^{m+1}},
\]
 as $|k |\to \iy$ uniformly in $\arg k \in [0,\pi]$, see \cite{BF60}, where
\[
\lb{1.18}
 Q_0={1\/2}\int_0^\iy q(x)dx,\ \ \ \   Q_1=-{i\/2^2}q(0), \
\ \ \ Q_2={1\/2^{3}}\big(q'(0)+\int_0^\iy  q^2(x)dx\big),...
\]
 In Theorem \ref{T3} we show that if $q\in W_{m+1}$ then the
function $M(k), k\in \C_+$ define by \er{Do2} satisfies
\[
\begin{aligned}
\lb{Kasm} M(k)={\cJ_0+iI_{0}\/t}+{\cJ_1+iI_{1}\/t^2}+...+
{\cJ_m+iI_{m}+o(1)\/t^{m+1}},
\end{aligned}
\]
 as $\Im k\to \iy$,\
where the real constsnts $I_j$ and $\cJ_j\in \R$ are given by
\[
\begin{aligned}
\lb{Kas}
  I_j=\Im Q_j,\qqq
\cJ_j={\rm v.p.} {1\/\pi}\int_\R h_{j-1}(t)dt, \qqq
h_{j}=t^{j+1}(h(t)+P_{j}(t)),\\
 h(t)=\log
|\p(t)|=h_{-1}(t),\qq
P_j(t)={I_{0}\/t}+{I_{1}\/t^2}+...+{I_{j}\/t^{j+1}},
\end{aligned}
\]
$j=0,1,2,...,m$. In particular, we have
\[
\lb{P0}
 \cJ_0={\rm v.p.}{1\/\pi}\int_\R h(t) dt,\qqq
 \cJ_1={\rm v.p.}{1\/\pi}\int_\R \big(th(t)+I_0\big) dt.
 \]
If $m=1$, then $\cJ_0$ in \er{P0} converges since
\er{apm} gives $h(t)=\Re{iQ_0\/t}+{O(1)\/t ^{2}}$ as $t\to \pm\iy$.

  \begin{theorem} \lb{T3} ({\bf Trace formulas})
Let a potential $q$ satisfy \er{dV}. Then
\[
\begin{aligned}
\lb{tr1} B_0+{\n(\R)\/\pi}+{1\/2}\int_0^\iy \Re q(x)dx=\cJ_0,
\end{aligned}
\]
where $\cJ_0$ converges. Let, in addition, a potential $q\in
W_{m+1}$ for some integer $m\ge 1$. Then the function $M$ satisfies
 \er{Kasm} and the following identities hold true
\[
\lb{trj} {B_j\/j+1}+K_j+\Re Q_j=\cJ_j,\qq j=1,...,m,
\]
in \ particular
\[
\begin{aligned}
{B_1\/2}+K_1+{\Im q(0)\/4}=\cJ_1,
\\
{B_2\/3}+K_2+\Re Q_2=\cJ_2.
\end{aligned}
\]

  \end{theorem}

\no {\bf Remark.} 1) Recall that  $B_0\ge 0$ and
$K_0={\n(\R)\/\pi}\ge 0$. Thus in order to estimate
$B_0+{\n(\R)\/\pi}\ge 0$ in terms of the potential $q$ we need to
estimate the integral $\cJ_0$ in terms of the potential $q$.

\medskip

Introduce the function $\f(r)={2\/\pi}\int_{0}^r {\sin t\/ t}dt,
r\ge 0$ and the integral $\mI$ by
\[
\lb{mJ} \mI={1\/2}\int_0^\iy \Re q(x)\f(2x\|q\|)dx.
\]
 Note that $\f(r)\in [0,1)$ for all $r\ge 0$ and $|\mI|\le
{1\/2}\int_0^\iy |\Re q(x)|dx$.

  \begin{theorem} ({\bf Estimates})
\lb{T4} Let a potential $q$ satisfy \er{dV}. Then the following
hold true
\[
\begin{aligned}
\lb{eBs} B_0+{\n(\R)\/\pi}+ \mI\le
{2\/\pi}\rt(\|q\|_1+\|q\|(C_0+\log^+ \|q\|)\rt),
\end{aligned}
\]
where $C_0=e^2+{e^4+1\/4}$  and
     $\log^+ z=\ca \log^+ z & z>1    \\
                           0& z\le 1 \ac$.
  \end{theorem}

\no {\bf Remark.} 1) We estimate the measure $\n(\R)$ plus sum of
$\Im k_j$ in terms of potentials.

2) If $\Re q=0$, then we have $\mI=0$. If $\pm \Re q\ge 0$, then we
have $\pm \mI\ge 0$.

3) In \cite{F18} there is an estimate of sum of $\d(E_j)$ in terms
of $\|q\|$, where $\d(z)=\dist (z,\R_+)$ for $z\in \C\sm [0,\iy)$.

Consider  estimates for  complex compactly supported potentials. In
this case the Jost function $\p(k)=f_+(0,k )$ is entire and due to
\er{1.2} it has a finite number of zeros in $\C_+$.

  \begin{theorem} \lb{Tcp}
Let $q\in L^1(\R_+)$ and let $\supp q\ss [0,\g]$ for some $\g>0$.
  Then  the number of zeros $\cN_+ (\p)$ of $\p$ (counted with
multiplicity) in $\ol\C_+$ satisfies
\[
\lb{qw2w} \cN_+ (\p) \le C_1  +C_2\g \|q\|,
\]
where the constants $C_1\sim 10,$ and $C_2\sim 1$ are given in Lemma
\ref{Tef}.

  \end{theorem}

Note that  the estimate of  $\cN_+$ was determined in \cite{FLS16},
when  $q$ decays exponentially at infinity.

\medskip

In our paper we use classical results from complex analysis about
the Hardy space in the upper half-plane. In particular, we use a
so-called canonical factorization of analytic functions from Hardy
spaced via its inner and outer factors, see Section 4. This gives us
to a new class of trace formula for the spectrum of Scr\"odinger
operators with complex-valued potentials on the half-line.

 We
shortly describe the plan of the paper. In Section~2 we present the
main properties of the Jost function. In Section~3 we prove main
theorems.  Section~4 is a collection of needed facts about Hardy
spaces. In Section~5 we discuss the case of compactly supported
potentials.

\section { Fundamental solutions}
\setcounter{equation}{0}

\medskip

\subsection{Fundamental solutions.}
It is well known that that the Jost solution $f_+(x,k )$ of equation
\er{1.3} satisfies the  integral equation
\[
\lb{2.1} f_+(x,k)=e^{ixk}+\int_x^\iy {\sin k(t-x)\/k}q(t)f_+(t,k)dt, \qqq  (x,k)\in [0,\iy)\ts \ol\C_+.
\]
In order to study properties of  the Jost function $\p(k)=f_+(0,k)$
we define the function $y(x,k)=e^{-ikx}f_+(x,k)$, which satisfies
the integral equation
\[
\lb{2.2} y(x,k)=1+\int _x^\iy G(t-x,k)q(t)y(t,k)dt, \qqq  G(t,k)={\sin
kt\/k}e^{ikt},
\]
$\forall\  (x,k)\in [0,\iy)\ts \ol\C_+.$ The standard iterations
give $y(x,k)$ and the Jost function $\p(k)=y(0,k)$:
\[
\lb{2.4}
\begin{aligned}
&y(x,k)=1+\sum_{n\ge 1}y_n(x,k),\\
& y_n(x,k)=\int _x^\iy G(t-x,k)q(t)y_{n-1}(t,k)dt, \qq y_0=1,\\
&\p(k)=1+\sum_{n\ge 1}\p_n(k),\qqq \p_n(k)=y_n(0,k),\\
& {\rm where} \qqq \p_1(k)=\int_0^\iy {\sin
kt\/k}e^{ikt}q(t)dt=-{q_0\/2ik}+{\wh q(k)\/2ik},
\end{aligned}
\]
and $\wh q(k)=\int _0^\iy e^{i2kt} \ q(t)dt$ is the Fourier
transformation. The identity \er{2.2} gives
\[
\begin{aligned}
\lb{2.3} \p(k)=y(0,k)=1+\int _0^\iy {\sin [kt]\/k} \ q(t)f_+(t,k)dt.
\end{aligned}
\]

We recall well-known properties of the functions introduced above
(see e.g., \cite{F59}).

\begin{lemma}
\label{TL2.1} Let $q\in L^1(\R_+)$ and let $x\in [0,\iy)$. Then the
functions $f_+(x,\cdot),f_+'(x,\cdot)$  are analytic in $\C_+$ and
continuous up to the real line without the point $0$.

Let, in addition, $\|q\|_1=\int_0^\iy x|q(x)|dx<\iy$. Then these
functions are continuous up to the real line.  Moreover, the
functions $y$ and $\p$ satisfy
\[
\lb{2.6} |y_n(x,k)|\le {\o^n\/n!},\qqq \o\in
\rt\{\|q\|_1,{\|q\|\/|k|}\rt\},\qqq n\ge 1,
\]
and
\[
\begin{aligned}
\lb{2.7}
|y(x,k)|\le e^{\o},\\
|y(x,k)-1|\le \o e^{\o},\\
|y(x,k)-1-y_1(x,k)|\le {\o^2\/2} e^{\o},
\end{aligned}
\]
and
\[
\lb{2.8}
\begin{aligned}
|\p(k)|\le  e^{\o},\qqq
 |\p(k)-1|\le \o e^{\o},\\
|\p(k)-1-\p_1(k)|\le {\o^2\/2} e^{\o}.
\end{aligned}
\]
In particular, if $\|q\|_1<\o_*$, where $\o_*e^{\o_*}=1$, then the
operator $H$ has not eigenvalues.

\end{lemma}
{\bf Proof.} Consider the first case, let $q\in L^1(\R_+)$ and let
$\o={\|q\|\/|k|}$. Substituting the estimate $ |G(t,k)|\le {1\/|k|}$
for all $t\ge 0, \ k\in \ol\C_+\sm\{0\}$ into the identity
\[
\lb{ynx}
y_n(x,k)=\int\limits_{x=t_0<t_1< t_2<...< t_n}
\lt(\prod\limits_{1\le j\le n}
G(t_{j}-t_{j-1},k)q(t_j)\rt)dt_1dt_2...dt_n,
\]
we obtain
\[
\begin{aligned}
\lb{2.10}
|y_n(x,k)|\le {1\/|k|^n}\int\limits_{x=t_0<t_1< t_2<...< t_n}
\lt(\prod\limits_{1\le j\le n}
|q(t_j)|\rt)dt_1dt_2...dt_n\\
= {1\/|k|^n}\int\limits_{x=t_0<t_1< t_2<...< t_n}
|q(t_1)q(t_2)....q(t_n)| dt_1dt_2...dt_n= {\|q\|^n\/n!
|k|^n}={\o^n\/n!}.
\end{aligned}
\]
This shows that for each $x\ge 0$ the series \er{2.4} converges
uniformly on bounded subset of $\ol\C_+\sm \{|k|>\ve\}$ for any
$\ve>0$. Each term of this series is an analytic function in $\C_+$.
Hence the sum is an analytic function in $\C_+$. Summing the
majorants we obtain estimates  \er{2.7}-\er{2.8} for $\o=\|q\|/|k|$.
Thus the functions $f_+(x,\cdot), f_+'(x,\cdot),x\in\R$ are analytic
in $\C_+$ and continuous up to the real line without the point $0$.

Consider the second case: let $\o=\|q\|_1=\int_0^\iy x|q(x)|dx<\iy$.
The function $G(t,k)$ satisfy $|G(t,k)|\le t$ for all $k\in \ol
\C_+, t>0$. Then using above arguments
 we obtain
\[
\begin{aligned}
\lb{2.15}
|y_n(x,k)|\le \int\limits_{x=t_0<t_1< t_2<...< t_n}
\lt(\prod\limits_{1\le j\le n}
|(t_{j}-t_{j-1})q(t_j)|\rt)dt_1dt_2...dt_n
\\
\le \int\limits_{x=t_0<t_1< t_2<...< t_n} \lt(\prod\limits_{1\le
j\le n} |t_{j}q(t_j)|\rt)dt_1dt_2...dt_n ={\|q\|_1^n\/n! }={\o^n\/n!
}.
\end{aligned}
\]
This shows that for each $x\ge 0$ the series \er{2.4} converges
uniformly on bounded subset of $\ol\C_+$. Each term of this series
is an analytic function in $\C_+$. Hence the sum is an analytic
function in $\C_+$. Summing the majorants we obtain estimates
\er{2.7}-\er{2.8} for $\o=\|q\|_1$. Thus the functions
$f_+(x,\cdot), f_+'(x,\cdot), x\ge 0$ are analytic in $\C_+$ and
continuous up to the real line.

We have $|f(k)-1|\le a e^{a}<1, a=\|q\|_1<\o_*$ for any $k\in
\ol\C_+$, which yields $|f(k)|\ge 1-a e^{a}>0$. \BBox

\section { Proof of main theorems }
\setcounter{equation}{0}

\subsection{Hardy spaces and Jost functions}

In order to study zeros of the Jost function $\p$ in the upper-half
plane we need to study the Blaschke product $B$, defined by \er{Bk}.
 We describe the basic properties
of the Blaschke product $B$ as an analytic function in $\C_+$.

{\bf  Proof Proposition \ref{T1}.} Lemma \ref{TL2.1} yields  that
the Jost function $\p$ is analytic in $\C_+$ and  is continuous  up
to the boundary  and satisfies \er{J1}. Moreover, the asymptotics
\er{2.7} implies that all zeros of $\p$ are uniformly bounded. Note
that (see page 53 in \cite{G81}), in general, in the upper half
plane the condition \er{B1} is replace by
\[
\lb{BLy} \sum {\Im k_j\/1+|k_j|^2}<\iy,
\]
and the Blaschke product with zeros $k_j$ has the form
\[
\lb{BL2x} B(k)={(k-i)^m\/(k+i)^m} \prod_{k_j\ne
0}^N{|1+k_j^2|\/1+k_j^2}\rt(\frac{k-k_j}{k-\ol k_j}\rt), \qqq k\in
\C_+.
\]
If all  moduli $|k_n|$ are uniformly bounded,  the estimate \er{BLy}
becomes $\sum \Im k_j<\iy$ and the convergence factors in \er{BL2x}
are not needed, since $\prod_{k_j\ne 0}^N \frac{k-k_j}{k-\ol k_j}$
already converges.

Results about zeros for the case $\|q\|_1<\o_*$ follows from Lemma
\ref{TL2.1}.

The statement i) is a standard fact for the function from
$\mH_\iy$, see Sect. VI  in \cite{Ko98}.

The statement ii) follows from Lemma \ref{TY2}. \BBox

\medskip

We describe the Jost function $\p(k), k\in \C_+$ in terms of a
canonical factorization.

{\bf Proof of Theorem \ref{T2}.} Lemma \ref{TL2.1}  gives  that the
Jost function $\p\in \mH_\iy$, $\p$ is continuous  in $\ol \C_+$ up
to the boundary and satisfies \er{2.8}. Thus from Theorem \ref{Tcf}
we obtain all results in Theorem \ref{T2}. \BBox

We prove the first result about the trace formulas.

{\bf  Proof of Corollary  \ref{T2x}.} Differentiating \er{Dio} and
using Theorem \ref{T2} we obtain
\[
\begin{aligned}
\lb{derfx}
{\p'(k)\/\p(k)}={B'(k)\/B(k)}-{i\/\pi}\int_\R{d\m(t)\/(k-t)^2},\qq
\qqq {B'(k)\/B(k)}=\sum {2i\Im k_j\/(k-k_j)(k-\ol k_j)},\qq \forall
\ k\in \C_+,
\end{aligned}
\]
where $d\m(t)=h(t)dt-d\n(t)$. Define $Y_0(k)=|q|^{1\/2}R_0(k)q_1,
k\in \C_+$, where $q_1=|q|^{1\/2}e^{i\arg V}$. Recall that $Y_0(k)$
is a trace class operator and the Jost function satisfies the
following identity $\p=\det (I+Y_0)$ in $\C_+$. The derivative of
the determinant $\p=\det (I+Y_0(k))$ satisfies
\[
\lb{derfz} {\p'(k)\/\p(k)}=-2k\Tr \big(R(k)-R_0(k)\big), \qq \forall
\ k\in \C_+,
\]
see \cite{GK69}. Combining \er{derfx}, \er{derfz} we obtain
\er{tre1}. Note that the series converges uniformly in every bounded
disc in $\C_+\sm \s_d$, since $\sum \Im k_j<\iy$.
 \BBox

\no {\bf  Proof of Theorem \ref{T3}.} Let a potential $q$ satisfy
\er{dV}. From Lemma \ref{TL2.1} and from asymptotics \er{2.4} we
deduce that  the Jost function $\p(k)=1-{1\/2ik}(q_0-\wh
q(k))+{O(1)\/k^2}$ as $k\to \pm \iy$. Thus from Theorem \ref{Tcftr}
we obtain \er{tr1}.

Let $q\in W_{m+1}, m\ge 1$. Then due to Lemma \ref{TL2.1},
asymptotics \er{apm} and  Theorem \ref{Tcftr},  we  obtain all
results in Theorem \ref{T3} for the case $m\ge 1$. \BBox

\no {\bf  Proof of Theorem \ref{T4}.} We estimate
$\cJ_0={1\/\pi}\int_0^\iy \x(t) dt$, where $
 \x(t)=\log |\p(t)\p(-t)|, t>0$ for the case $r_c=\|q\|>1$, the proof for
 the case $r_c<1$ is similar.
 We rewrite $\cJ_0$ in the following form
\[
\lb{F1}
\begin{aligned}
\cJ_0 =\cJ_{01}+\cJ_{02},\qqq \cJ_{01}={1\/\pi}\int_0^{r_c} \x(t)
dt,\qqq \cJ_{02}={1\/\pi}\int_{r_c}^\iy \x(t) dt.
 \end{aligned}
 \]
Consider $\cJ_{01}$. The estimate \er{2.8} with $\o={\|q\|_1}$ for
the intreval $(0,1)$ and with $\o={\|q\|\/|k|}$ for $(1,r_c)$ gives
\[
\lb{J01}
\begin{aligned}
\cJ_{01}={1\/\pi}\int_0^1\x(t)dt +{1\/\pi}\int_1^{r_c}\x(t) dt \\
\le {2\/\pi}\int_0^1 \|q\|_1 dt+
{2\|q\|\/\pi}\int_1^{r_c} {dt\/t}={2\|q\|_1\/\pi}+{2\|q\|\/\pi}\log r_c.
 \end{aligned}
\]

Consider $\cJ_{02}$. The estimate \er{2.6} gives
\[
\begin{aligned}
\lb{F1x}
\p(k)=1+g(k),\qqq g(k)=\p_1(k)+\wt \p(k),\\
|\p(k)|\le e^{\o},\qqq |\p(k)-1|\le \o e^{\o},\qqq |\wt \p(k)|\le
{\o^2\/2} e^{\o},
\end{aligned}
 \]
where
\[
\lb{axxx} \o={\|q\|\/|k|},\qqq \qq \p_1(k)=\int_0^\iy{\sin
kx\/k}e^{ikx}q(x)dx.
\]
Let $F^-(k)=F(-k)$. This and the identities
$g+g^-=(\p_1+\p_1^-)+(\wt \p+\wt \p^-)=f_0+(\wt \p+\wt \p^-)$ yield
\[
\lb{asxd}
\begin{aligned}
& f=\p\p^-=(1+g)(1+g^-)=1+g+g^-+gg^-=1+f_0+f_1,\\
& f_0(k):=\p_1(k)+\p_1(-k)={1\/k}\int_0^\iy q(x)\sin [2kx] dx,
\\
& f_1:=\wt \p+\wt \p^-+gg^-.
\end{aligned}
\]
 Then the estimates \er{F1x} implies
\[
\lb{asp1Vxx}
\begin{aligned}
& |\wt \p(k)+\wt \p(-k)|\le {\o^2} e^{2\o},\qqq |g(k)g(-k)|\le
{\o^2}e^{2\o},
\end{aligned}
\]
which yields
\[
\lb{ah1}
\begin{aligned}
f=1+f_0+f_1,\qqq |f_0|\le \o,\qq |f_1(k)|\le 2{\o^2} e^{2\o}.
\end{aligned}
\]
This yields
\[
\lb{ah1x}
\begin{aligned}
f\ol f=(1+f_0+f_1)(1+\ol f_0+\ol f_1)=1+2\Re (f_0+f_1)+|f_0+f_1|^2
\end{aligned}
\]
and then
\[
\begin{aligned}
\x={1\/2}\log |f|^2\le {1\/2}(2\Re f_0+F), \qq F=2 \Re
f_1+|f_0+f_1|^2.
\end{aligned}
\]
Thus we obtain
\[
\lb{J2}
\begin{aligned}
\cJ_{02}={1\/2\pi}\int_{r_c}^\iy \log |f^2(k)| dk\le
{1\/2\pi}\int_{r_c}^\iy (2\Re f_0+F)dk.
\end{aligned}
\]
Here due to \er{asxd} the first integral has the form
\[
\lb{J2x}
\begin{aligned}
{1\/\pi}\int_{r_c}^\iy \Re f_0dk ={1\/2}\int_0^\iy \Re
q(x)\vp(x)dx={1\/2} \int_0^\iy \Re q(x)(1-\f(r)dx,\qq r=2xr_c,
\end{aligned}
\]
where $\f(r)={2\/\pi}\int_{0}^r {\sin t\/ t}dt$ and $\f(r)\in
[0,1]$, since the function $\vp(x)={2\/\pi} \int_{r_c}^\iy {\sin
2kx\/ k}dk$ satisfies:
\[
\begin{aligned}
& \vp(x)={2\/\pi} \int_{r_c}^\iy {\sin 2kx\/
k}dk={2\/\pi}\int_{r}^\iy {\sin t\/ t}dt=\vp(0)-\f(r)=1-\f(r).
\end{aligned}
\]
Recall that we take  $\o={r_c\/k}$. We estimate the second integral:
\[
\lb{eF}
\begin{aligned}
& F=2 \Re f_1+ |f_0 +f_1(k)|^2\le 4{\o^2} e^{2\o}+(\o+2{\o^2}
e^{2\o})^2 \le 8{\o^2} e^{2\o}+ \o^2  +4{\o^2} e^{4\o},
\\
& \int_{r_c}^\iy {\o^2} e^{A\o}dt= \int_{r_c}^\iy {r_c^2\/t^2}
e^{Ar_c\/t}dt=-{r_c\/A}e^{Ar_c\/t}\rt|_{r_c}^\iy ={r_c\/A}e^A, \ \
\forall \ A\ne 0,
\\
& {1\/2\pi}\int_{r_c}^\iy Fdk\le {1\/2\pi}\int_{r_c}^\iy(8{\o^2}
e^{2\o}+ \o^2+4{\o^2} e^{4\o})dk={r_c\/2\pi}(4e^2+1+
e^4)={2r_c\/\pi}C_0,
\end{aligned}
\]
where $C_0={4e^2+1+ e^4\/4}$. Substituting  \er{J01}, \er{J2},
\er{J2x}, \er{eF} into \er{tr1} we obtain
$$
\begin{aligned}
B_0+{\n(\R)\/\pi}+{1\/2}\int_0^\iy \Re q(x)dx=\cJ_0
\\
\le {2\/\pi}\rt(\|q\|_1+\|q\|\log r_c\rt)+ \rt({1\/2} \int_0^\iy \Re
q(x)(1-\f(r)dx+
       {2r_c\/\pi}C_0 \rt),
\end{aligned}
$$
which yields \er{eBs}. \BBox



\

\section {Analytic functions in the upper half-plane}
\setcounter{equation}{0}

\

We discuss different properties of functions from Hardy spaces.
Recall that if $f\in \mH_\iy(\C_+)$, then the Blaschke product $B\in
\mH_\iy$ with $\|B\|_{\mH_\iy}\le 1$ and
\[
\lb{B3} \lim_{v\to+0} B(u+iv)=B(u+i0), \qqq |B(u+i0)|=1 \qqq   \
{\rm a. e \ on} \ \R,
\]
\[
\lb{B4} \lim _{v\to 0}\int_\R\log |B(u+iv)|du=0.
\]
We recall the needed results about the Blaschke product (see e.g.
\cite{K17}).

\begin{lemma}
\lb{TY2} Let $f\in \mH_\iy(\C_+)$ and let $\{z_j\}$ be the zeros of
$f$ in $\C_+$, which are uniformly bounded by $r_0$. Define $
B_n=2\sum_{j} \Im z_j^{n+1}$ for all $n\ge 0$. Then
\[
\lb{AB2} |B_n|\le 2\sum_{j} |\Im z_j^{n+1}|\le
{\pi\/2}(n+1)r_0^{n}B_0 \qqq \qqq \forall \ n\ge 1.
\]
Moreover, the function $\log B(z)$ is analytic in $\{|z|>r_0\}$ and
has the corresponding Taylor series given by
\[
\lb{AB3} \log
B(z)=-{iB_0\/z}-{iB_1\/2z^2}-{iB_2\/3z^3}-....-{iB_{n-1}\/nz^n}-....
\]

\end{lemma}

We need some results about functions from Hardy spaces. We begin
with asymptotics. Consider the integral
\[
M(k)={1\/\pi}\int_\R { h(t)\/k-t}dt,\qqq k\in \C_+,
\]
where $h$ belongs to  a class $\gX_m $ defined by

{\bf Definition.} {\it A function $h$ belongs to the class
$\gX_m=\gX_m(\R)$ if $h\in L_{real, loc}^1(\R)$ and has the form
\[
\lb{M1}
\begin{aligned}
 h(t)=-P_m(t)+{h_m(t)\/t^{m+1}},\qqq
P_m(t)={I_{0}\/t}+{I_{1}\/t^2}+...+{I_{m}\/t^{m+1}},
\\
h_m(t)=o(1)\qqq as \qqq t\to \pm \iy
\end{aligned}
\]
for some real constants $I_0,I_1,....,I_m$ and integer $m\ge 0$ and
there exist following integrals}
\[
\lb{M2}
\begin{aligned}
\cJ_m={\rm v.p.}{1\/\pi}\int_\R h_{m-1}(t)dt,\qqq\qqq M_m(k)={\rm
v.p.}{1\/\pi}\int_\R { h_m(t)\/k-t}dt, \qqq  k\in \C_+.
\end{aligned}
\]

\

Note that if $h\in \gX_m $ for some $m\ge 0$, then there exist
finite integrals (the principal value):
\[
\begin{aligned}
\lb{M3}  \cJ_j={1\/\pi}\lim_{s\to \iy}\int_{-s}^s h_{j-1}(t)dt, \qqq
h_{j}=t^{j+1}(h(t)+P_{j}(t)),\qq h_{-1}=h,
\end{aligned}
\]
for all $j=0,1,2,...,m-1$.

\begin{lemma}
\lb{Th1} i) Let $M(k)={1\/\pi}\int_\R { h(t)\/k-t}dt$  for some
$h\in \gX_m, m\ge 0$. Then
\[
\begin{aligned}
\lb{M4} M(k)={\cJ_0+iI_{0}\/k}+{\cJ_1+iI_{1}\/k^2}+...
+{\cJ_m+iI_{m}\/k^{m+1}}+{M_m(k)\/k^{m+1}}, \qq \forall \ k\in \C_+,
\end{aligned}
\]
 where the constants $\cJ_0,....,\cJ_{m} \in \R$ and the function
$M_m$ are given by \er{M2}-\er{M3}.

ii) Let $h$ be a real function from  $L_{loc}^1(\R) $ and satisfy
\er{M1} for some $m\ge 0$ and
\[
\lb{M6}
\begin{aligned}
h_m=\a+\Re\wh \b,\qq  \where \ \ \a\in L^s(\R)\cap
L^\iy(\R),\qq s\in [1,\iy),\\
\wh \b(t)=\int_0^\iy e^{i2tx}\b(x)dx,\qqq
\b\in L^1(\R).
\end{aligned}
\]
 Then $h\in \gX_m$ and
the function $M_m$ from \er{M2} is analytic in $\C_+$ and
\[
\lb{Mm} M_m(k)=o(1)\qqq \as \qq \Im k\to \iy.
\]

\end{lemma}

\no {\bf  Proof.} The statement i) was proved in \cite{K17x}.

ii) We consider $M_m$ in \er{M2}. It is clear that the function $\wt
\a(k)={1\/\pi}\int_\R { \a(t)\/k-t}dt$ is well defined and is
analytic in $\C_+$ and $\wt \a(k)=o(1)$ as $\Im k\to\pm\iy$.

In the case $\b=\b_1+i\b_2$ we use the following identities
$$
\Re\wh \b(t)=\int_0^\iy \Re e^{i2tx}\b(x)dx, \qq \Re e^{i2tx}\b=\Re
(c+is)(\b_1+i\b_2)=c\b_1-s\b_2,
$$
$$
\int_\R { \hat g(t)\/k-t}dt=-2\pi i\int_0^\iy e^{ikx}g(x)dx, \qqq
\int_\R { \hat g(-t)\/k-t}dt=0, \qq \forall\ k\in \C_+,
$$
for $g\in L^1(\R_+)$ and where $c=\cos 2tx, s=\sin 2tx$. These
identities yield
$$
{1\/\pi}\int_\R { \Re \hat \b(t)\/k-t}dt=-{i\/2}\hat \b(k), \qq
\forall\ k\in \C_+.
$$
Thus the function $M_m$ from \er{M2} is analytic in $\C_+$ and $
M_m(k)=o(1)$ as $\Im k\to\pm\iy$.

The proof  of the existence of $\cJ_m={\rm v.p.}{1\/\pi}\int_\R
h_{m-1}(t)dt$ in \er{M2} is similar, since $h_{m-1}={1\/t}(-I_m+h_m)
$ and $h_m=\a+\Re\wh \b$ and $h_{m-1}, h_m\in L^1_{loc}(\R)$. \BBox

We recall the standard facts about the canonical factorization, see
e.g. \cite{G81}, \cite{Ko98} and  in  the needed form for us from \cite{K17x}.

\begin{theorem}\label{Tcf}
Let a function $f\in\mH_p$ for some $p\ge 1$ and be a continuous in
$\ol \C_+$ and $f(k)=1+O(1/k)$ as $|k|\to \iy$, uniformly in
${\arg}\,k \in [0,\pi]$.
 Then $f(k),  k\in \C_+$ has a canonical factorization in $\C_+$ given by
\[
\begin{aligned}
\lb{af1} & f=f_{in}f_{out}, \qqq  f_{in}(k)=B(k)e^{-iK(k)},\qqq
K(k)={1\/\pi}\int_\R {d\n(t)\/k-t},
\\
& f_{out}(k)=e^{iM(k)},\qqq  M(k)= {1\/\pi}\int_\R {\log
|f(t)|\/k-t} dt.
\end{aligned}
\]
$\bu$ $B$ is the Blaschke product for $\Im k>0$ given by \er{Bk} and
 $d\n(t)\ge 0$ is some singular compactly supported measure on
$\R$ and for some $r_c>0$ satisfies
\[
\lb{sms}
\begin{aligned}
\n(\R)=\int_\R d\n(t)<\iy,\qqq
 \supp \n \ss [-r_c, r_c],
 \end{aligned}
\]
\[
\lb{ms}
 \supp \n \ss \{k\in \R: f(k)=0\}\ss [-r_c, r_c].
\]
$\bu$  The function $K(\cdot)$ has an analytic continuation from
$\C_+$ into the domain $\C\sm [-r_c, r_c]$   and has the following
Taylor series
\[
\lb{Knx} K(k)=\sum_{j=0}^\iy {K_j\/k^{j+1}},\qqq K_j={1\/\pi}\int_\R
t^jd\n(t).
\]
$\bu$ The function $\log |f(t+i0)|$ belongs to  $L_{loc}^1(\R)$.

\end{theorem}

{\bf Remark.}  The integral $M(k)$ in \er{af1} converges absolutely
since $f(t)=1+{O(1)\/t}$ as $t\to \pm\iy$.

\medskip

In order to describe the Jost function $\p$ in terms of a canonical
factorization we introduce the corresponding class of  functions.
Let $f\in \mH_p$ for some $0<p\le \iy$ and for integer $m\ge 0$  $f$
satisfies
\[
\lb{afm}
  -i\log f(k)={Q_0\/k }+{Q_1\/k ^{2}}+{Q_2\/k ^{3}}+\dots
  +{Q_m+o(1)\/k ^{m+1}},
\]
 as $|k |\to \iy$ uniformly in $\arg k \in [0,\pi]$, for some
constants $Q_j\in \C, j=0,1,2...,m$. Then the function
$h=\log|f(t)|, t\in \R$ has the form
\[
\lb{hx}
\begin{aligned}
 h(t)=-P_m(t)+{h_m(t)\/t^{m+1}},\qqq
P_m(t)={I_{0}\/t}+{I_{1}\/t^2}+...+{I_{m}\/t^{m+1}},
\\
I_j=\Im Q_j,\ j=0,1,..,m,\qqq h_m(t)=o(1)\qqq as \qqq t\to \pm \iy.
\end{aligned}
\]

We describe a canonical factorization of functions from $\mH_p^m$
from \cite{K17x}.

\begin{theorem}\label{Tcftr}
Let a function $f\in \mH_p$  satisfy \er{afm} for some $m\ge 0, p\ge
1$ and let the function $h(t)=\log |f(t)|, t\in \R$ has the form
\er{hx}, where $h_m$ satisfies \er{M6}. Then  $f$ has a canonical
factorization $ f=f_{in}f_{out}$ in $\C_+$ given by Theorem
\ref{Tcf}, where  the function $M$ has the form
\[
\begin{aligned}
\lb{aM} M(k)={1\/\pi}\int_\R {
h(t)\/k-t}dt={\cJ_0+iI_{0}\/k}+{\cJ_1+iI_{1}\/k^2}+...
+{\cJ_m+iI_{m}+M_m(k)\/k^{m+1}},\\
\end{aligned}
\]
for any $k\in \C_+$, where
\[
\begin{aligned}
\lb{aMx} M_m(k)={1\/\pi}\int_\R { h_m(t)\/k-t}dt,\qq
 \cJ_j=v.p.{1\/\pi}\int_\R h_{j-1}(t)dt,
\qqq h_j(t)=th_{j-1}-I_j,
\end{aligned}
\]
\[
\lb{aM3} M_m(k)=o(1)\qq \as \qq \Im k\to \iy,
\]
$j=0,1,...,m,$ and  $h_{-1}=h$. Moreover, the following trace
formulas hold true:
\[
\lb{atr} B_j+K_j+\Re Q_j=\cJ_j, \qqq j=0,1,..,m.
\]
\end{theorem}

\section {Schr\"odinger operators with compactly supported potentials}
\setcounter{equation}{0}

\subsection{Entire functions}
An entire function $f(k)$ is said to be of exponential type if there
is a constant $\b$ such that $|f(k)|\leq\const e^{\b |k|}$
everywhere. The infimum of the set of $\b$ for which such inequality
holds is called the type of $f$.

\no {\bf Definition.} {\it Let $\cE_\g, \g>0$ denote the class of
exponential type  functions $f$ satisfying
\[
\lb{dE}
\begin{aligned}
&|f(k)-1|\le \o(k) e^{2\g k_-+\o(k) } \qqq
\forall \ k\in \C,\ \\
 & k_-={1\/2}(|\Im k|-\Im k)\ge 0,\qqq \o(k)=\min \{Q_1, {Q\/|k|} \},\qqq
  \qqq
\end{aligned}
\]
 for some positive constants $Q_1, Q$ and if
 each its zero $z$ in $\ol\C_+$ satisfies $|z|\le Q$.
}

Note that if $q\in L^1(\R_+)$ and $\supp q\ss [0,\g]$, then a Jost
function $\p\in \cE_\g$, see below the proof of Theorem \ref{Tcp}.
Define the disk $\dD_r(u)=\{z: |z-u|<r\}$ for $u\in \C$ and $r>0$.

\begin{lemma}
\label{Tef} i)  Let $f\in \cE_\g$ and let $Q_1<\o_*$, where
$\o_*e^{\o_*}=1$. Then $f(k)\ne 0$ for any $k\in \ol\C_+$.

ii)  Let $f\in \cE_\g$  and let $\o(r)\le {1\/2}$ for some $r>0$. If
$k\in \C_+ ,|k|\le r$, then
\[
\lb{k0} |f(k)|\ge 1-{\sqrt e\/2}.
\]
Moreover,  the number of zeros $\cN(\r)$ of $f$ (counted with
multiplicity) in disk $\dD_\r(it)$ with the center $it=i2Q$ and the
radius $\r=\sqrt\a  Q $ for any $\a\in (5,8)$ satisfies
\[
\lb{qw2} \cN(\r) \le C_1+C_2 \g Q,
\]
where  the constants $C_1={1\/C_*}\rt({\sqrt2+2\/4}+ \log
{2\/2-\sqrt 2} \rt)$ and $C_2={4-\pi\/ \pi C_*}$ and $C_*={1\/2}\log
{8\/\a}>0$.

In particular, if we take $\a-5$ small enough,  then $C_1\sim 10$
and $C_1\sim 1$, and the domain $\{k\in \ol\C_+, |k|\le Q\}\ss
\dD_\r(it)$.

\end{lemma}

\no {\bf Proof.} i) We have $|f(k)-1|\le Q_1 e^{Q_1}<1$ for any
$k\in \ol\C_+$, which yields $|f(k)|\ge 1-Q_1 e^{Q_1}>0$.

ii) From \er{dE} and $\o(r)\le {1\/2}$ we obtain $ |f(k)-1|\le \o(r)
e^{\o(r)}\le {1\/2}e^{{1\/2}}, $ which yields \er{k0}.

 Recall the Jensen formula (see p. 2 in \cite{Koo88}) for an entire
function $F$ any $r>0$:
\[
\lb{qw3} \log |F(0)|+\int _0^r{\cN_s (F)\/s}ds= {1\/ 2\pi }\int
_0^{2\pi}\log |F(re^{i\f})|d\f,
\]
where $ \cN_s (F)$ is the number of zeros of  $F$ in the disk
$\dD_s(0)$. We take the function $F(z)=f(it+z)$ and the disk
$\dD_r(it), it=i2Q$ with the radius $r=\sqrt 8Q$. Thus \er{qw3}
implies
\[
\begin{aligned}
\lb{qw5} \log |f(it)|+\int _0^r{\cN (s)\/ s}ds =S, \qqq S={1\/ 2\pi
}\int _0^{2\pi}\log \big|f(k_\f)\big|d\f,\qq k_\f=it-ire^{i\f},
\end{aligned}
\]
where $\cN (s)=\cN_s (f(it+\cdot))$.
 From \er{dE} we obtain at $k_\f=it-ire^{i\f}$ and $a={\pi\/4}$:
\[
\lb{qw7}
\begin{aligned}
S={1\/ 2\pi }\int _0^{2\pi}\log \big|f(k_\f)\big|d\f=S_0+S_1, \qqq
S_0={1\/ 2\pi }\int _{ -a}^{a}\log \big|f(k_\f)\big|d\f,
\\
S_1={1\/ 2\pi }\int _a^{2\pi -a}\log \big|f(k_\f)\big|d\f\le {1\/
2\pi }\int _a^{2\pi-a}\log \big(1+\o_0 e^{\o_0}\big)d\f={3 \/4}\log
\big(1+\o_0 e^{\o_0}\big),
\end{aligned}
\]
where $\o_0=\o(2Q)\le {1\/2}$. On the circle $|it-iz|=r$ using
\er{dE} we get
\[
\lb{qw10}
\begin{aligned}
& \min_{0\le\f\le a} |it-ire^{i\f}|=(\sqrt8-2)Q ,
\\
& \o_1= \max_{0\le\f\le a}\o(it-ire^{i\f})=
{1\/\sqrt8-2}={\sqrt2+1\/2}, \qq \log \big(1+\o_1e^{ \o_1 }\big)\le
2\o_1.
\end{aligned}
\]
Define the integral $S_{00}={2\g\/ \pi }\int _{0}^{a} k_- d\f$,
where we have $k_-=r\sin \f -t=r(\sin \f-\sin a)$ for $\f\in (-a,a)$
and \er{dE}, \er{qw10} give
\[
\lb{qw8}
\begin{aligned}
S_0={1\/ 2\pi }\int _{ -a}^{a}\log \big|f(k_\f)\big|d\f \le {1\/ \pi
}\int _{0}^{a}\log \big(1+\o_1 e^{2\g k_-+\o_1 }\big)d\f
\\
\le S_{00}+ {1\/ \pi }\int _{0}^{a} \log \big(1+\o_1e^{ \o_1
}\big)\f=S_{00}+{ \o_1\/2}\le S_{00}+{\log \big(1+\o_1e^{ \o_1
}\big)\/4},
\end{aligned}
\]
and
\[
\lb{qw9}
\begin{aligned}
S_{00}= {2\g\/ \pi }\int _{0}^{a} k_-d\f= {2\g\/ \pi }\int _{0}^{a}
r(\cos \f-\cos a)   d\f={2 \g r\/ \pi } (\sin a-a\cos a)
=\g Q{(4-\pi)\/ \pi }.
\end{aligned}
\]
Collecting \er{qw7}-\er{qw9} we obtain
\[
\lb{qw11} S\le {1\/4}+ {\sqrt2+1\/4} + \g Q{(4-\pi)\/ \pi }=
{\sqrt2+2\/4}+\g Q{(4-\pi)\/ \pi }.
\]
Thus if $\r=\sqrt \a Q$ for any $\a\in (5,8)$, then we get $
\dD_{\r}(it)\ss \dD_{r}(it)$, which yields
\[
\lb{qw12}
\begin{aligned}
 \int _0^r{\cN (t)dt\/ t}\geq \cN (\r)\int _{\r}^r{dt\/ t}=\cN
(\r)\log {r\/\r}=C_*\cN (\r), \qq C_*={1\/2}\log {8\/\a}>0.
\end{aligned}
\]
 Then substituting  \er{k0}, \er{qw11},
\er{qw12} into \er{qw5} we obtain
\[
\lb{qw13}
\begin{aligned}
\log {2-\sqrt e\/2}+C_*\cN (\r) \le \log |f(it)|+C_*\cN (\r) \le
{\sqrt2+2\/4}+\g Q{(4-\pi)\/ \pi },
\end{aligned}
\]
which yields \er{qw2}. If $\a-5$ is small enough,  then $C_1\sim 10$
and $C_1\sim 1$, and the domain $\{\Im k\ge 0, |k|\le Q\}\ss
\dD_\r(it)$.   \BBox


\subsection{Proof of Theorem \ref{Tcp}}
Consider  the Schr\"odinger operator $H$, when the potential $q\in
L^1(\R_+)$  is  complex and
 $\supp q\ss [0,\g]$ for some $\g>0$. We recall known results: {\it  The Jost function
 $\p$  is entire and satisfies
$
 |\p(k)-1|\le \o(k) e^{2\g k_- +\o(k)}
$ for all $k\in \C$, where $\o(k)=\min \{\|q\|_1, {\|q\|\/|k|_1}\}$
(see e.g., \cite{K16}).}  Thus due to this fact and \er{1.2} we
obtain that the function $\p\in \cE$ with $Q=\|q\|$ and
$Q_1=\|q\|_1$. Then Lemma \ref{Tef} gives the estimate \er{qw2w}.
 \BBox

\medskip


\footnotesize

\no {\bf Acknowledgments.} \footnotesize Evgeny Korotyaev is
grateful to Ari Laptev for discussions about the Schr\"odinger
operators with complex potentials.
 He is also grateful to Alexei Alexandrov (St. Petersburg) for
discussions and useful comments about Hardy spaces. Our study was
supported by the RSF grant No 18-11-00032.

\end{document}